\documentclass[aps,prl,preprint,superscriptaddress,showpacs]{revtex4}
\usepackage{graphicx}
\usepackage{dcolumn}
\usepackage{bm}
\usepackage[usenames]{color}
\usepackage{times}
\newcommand{\comment}[1]{}
\def\vs{\textit{vs.}}

\def\EF{$E_{\mathrm{F}}$}
\def\ED{$E_{\mathrm{D}}$}

\def\CaC6{CaC$_{6}$}
\def\C6{C$_{6}$}
\def\KC8{KC$_{8}$}
\def\6r3{$6\sqrt{3}\times 6\sqrt{3}$}

\def\spthree{$sp^{3}$}
\def\pz{$p_{z}$}
\def\nh{$n_{\mathrm{H}}$}
\def\nk{$n_{\mathrm{K}}$}

\def\kf{$k_{\mathrm{F}}$}

\def\2x2{$(2 \times 2)$}
\def\1x1{$(1 \times 1)$}
\def\RT3{$\sqrt{3}\times\sqrt{3}$-R30}
\def\rt3x{$\sqrt{3}\times$}

\def\dk{$\Delta k$}

\def\dw{$\mathcal{D}(\omega)$}

\def\A1{\AA$^{-1}$}

\def\kF{$k_{F}$}

\def\pistar{$\pi^{*}$}

\definecolor{eli}{cmyk}{1.0, 0.0, 1.0, 0.65} 
\definecolor{suppl}{cmyk}{1.0, 1.0, 0.0, 0.35}

\def\lmfp{$L_{\mathrm{mfp}}$}
\def\lh{$L_{\mathrm{H}}$}

\begin{document}

\preprint{APS/123-QED}

\title{Quasiparticle Transformation During a Metal-Insulator Transition in Graphene}

\author{Aaron Bostwick}
    \affiliation{Advanced Light Source, Lawrence Berkeley National
    Laboratory, Berkeley, CA 94720, USA}
    
\author{Jessica L. McChesney}
    \affiliation{Advanced Light Source, Lawrence Berkeley National
    Laboratory, Berkeley, CA 94720, USA}

\author{Konstantin Emtsev}
    \affiliation{Lehrstuhl f\"ur Technische Physik, Universit{\"a}t Erlangen-N{\"u}rnberg, 91058 Erlangen, Germany}

\author{Thomas Seyller}
    \affiliation{Lehrstuhl f\"ur Technische Physik, Universit{\"a}t Erlangen-N{\"u}rnberg, 91058 Erlangen, Germany}

\author{Karsten Horn}
    \affiliation{Department of Molecular Physics, Fritz-Haber-Institut der Max-Planck-Gesellschaft, 14195 Berlin, Germany}

\author{Stephan D. Kevan}
\affiliation{Department of Physics, University of Oregon, Eugene OR 97403}
    
\author{Eli Rotenberg}
    \affiliation{Advanced Light Source, Lawrence Berkeley National Laboratory, Berkeley, CA 94720, USA}

\date{\today}


\def\figonecaption{Fermi surfaces (upper) and associated bandstructure cuts (lower) through the graphene K point for (a) clean, and
(b-d) as a function of \nh\ indicated in H atoms per cm$^{2}$.  The dashed lines schematically show the Fermi surface contours.  The
solid lines in the lower row are the peak position determined by fits to the momentum distribution curves.}

\def\figone{\begin{figure*} \begin{center}
    {\includegraphics[width=6.250in]{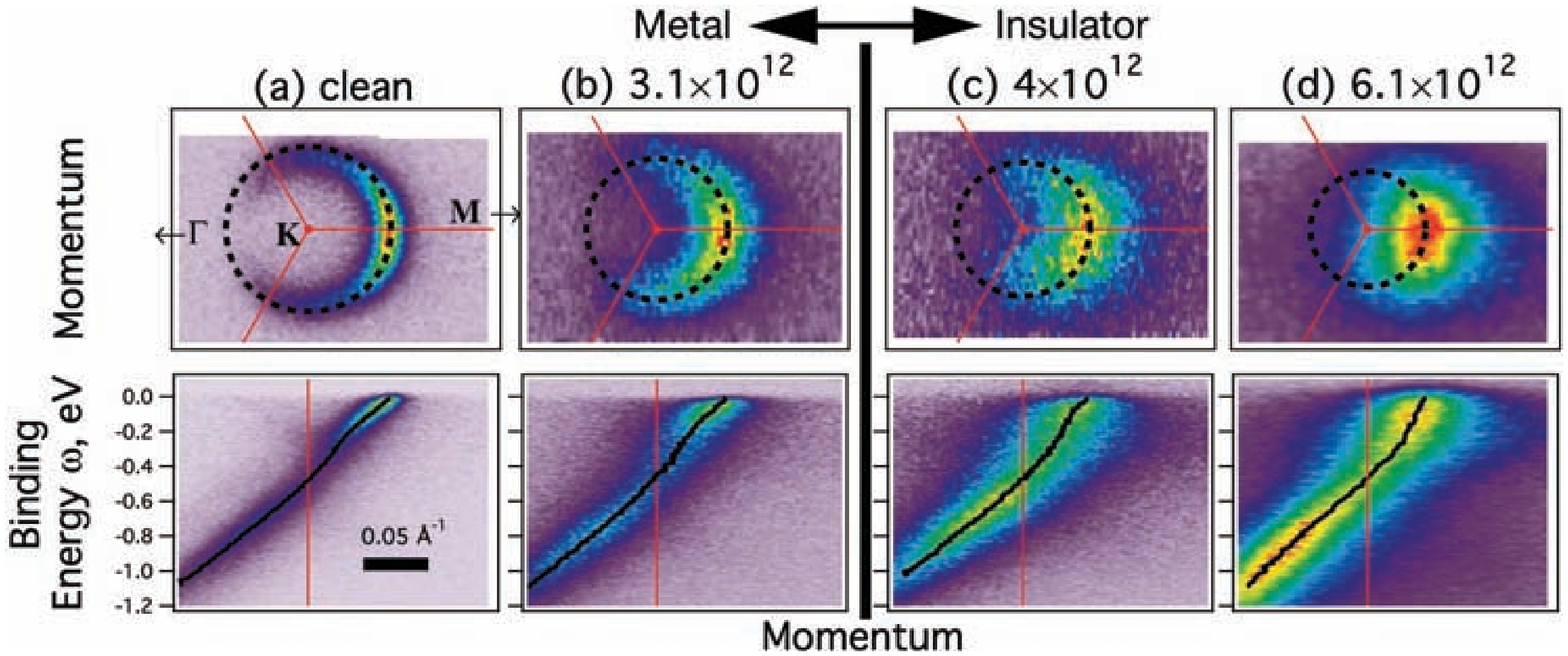}}
    \caption{\figonecaption\label{fig:1}}
\end{center}
\end{figure*}
}
\def\figoneref{Fig.\ \ref{fig:1}}

\def\figtwocaption{(a-d) Variation of (a) DOS, fitted to lines, (b) momentum linewidth, (c) Energy distribution curves (EDCs) at
\kf\ and (d) EDCs at the K symmetry point.  The measurements were conducted under constant dosing conditions for the hydrogen
coverages ($\times 10^{12}$ cm$^{-2}$) indicated in (d) with each trace averaged over 4 spectra at slightly different coverages.
The horizontal lines indicate the baselines for each trace which are shifted for clarity.  (e) Angle-integrated spectra for clean
and dosed graphene, together with the difference spectrum, acquired $\sim 0.3\mathrm{\AA}^{-1}$ away from the K point.  }

\def\figtwo{\begin{figure*} \begin{center}
{\includegraphics[width=7in]{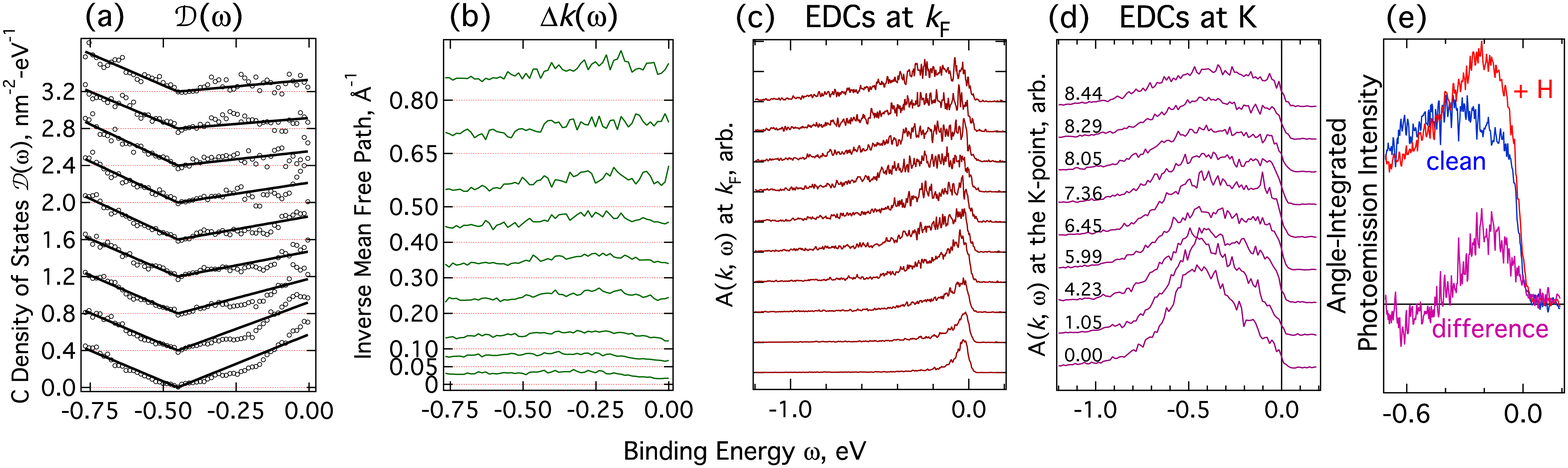}}
    \caption{\figtwocaption\label{fig:2}}

\end{center}
\end{figure*}
}
\def\figoneref{Fig.\ \ref{fig:1}}

\def\figtworef{Fig.\ \ref{fig:2}}

\def\figthreecaption{(a) The momentum distribution at \EF\ as a function of coverage \nh.  (b) The momentum width $\Delta
k$ as a function coverage \nh\ and \nk\ for atomic H or K, resp., compared to the resistance $R$ determined by photocurrent-induced
voltage change as a function of \nh.  The data were taken simultaneously with the data in Fig.\ 2.  The error bars for $R$ are the
same or smaller than the symbols.  (c) The relevant length scales as a function of atomic H coverage \nh: the defect length scale
\lh, the inverse momentum width 1/$\Delta k$, and the inverse Fermi wavevector 1/\kf.  The crossing of the latter two curves defines
the Ioffe, Regel, and Mott (IR\&M) criterion for the localization transition; the observed MIT occurs at the coverage indicated by
the gold line.  The red and blue lines are guides to the eye.} \def\figthree{\begin{figure}\begin{center}
{\includegraphics[width=6.25in]{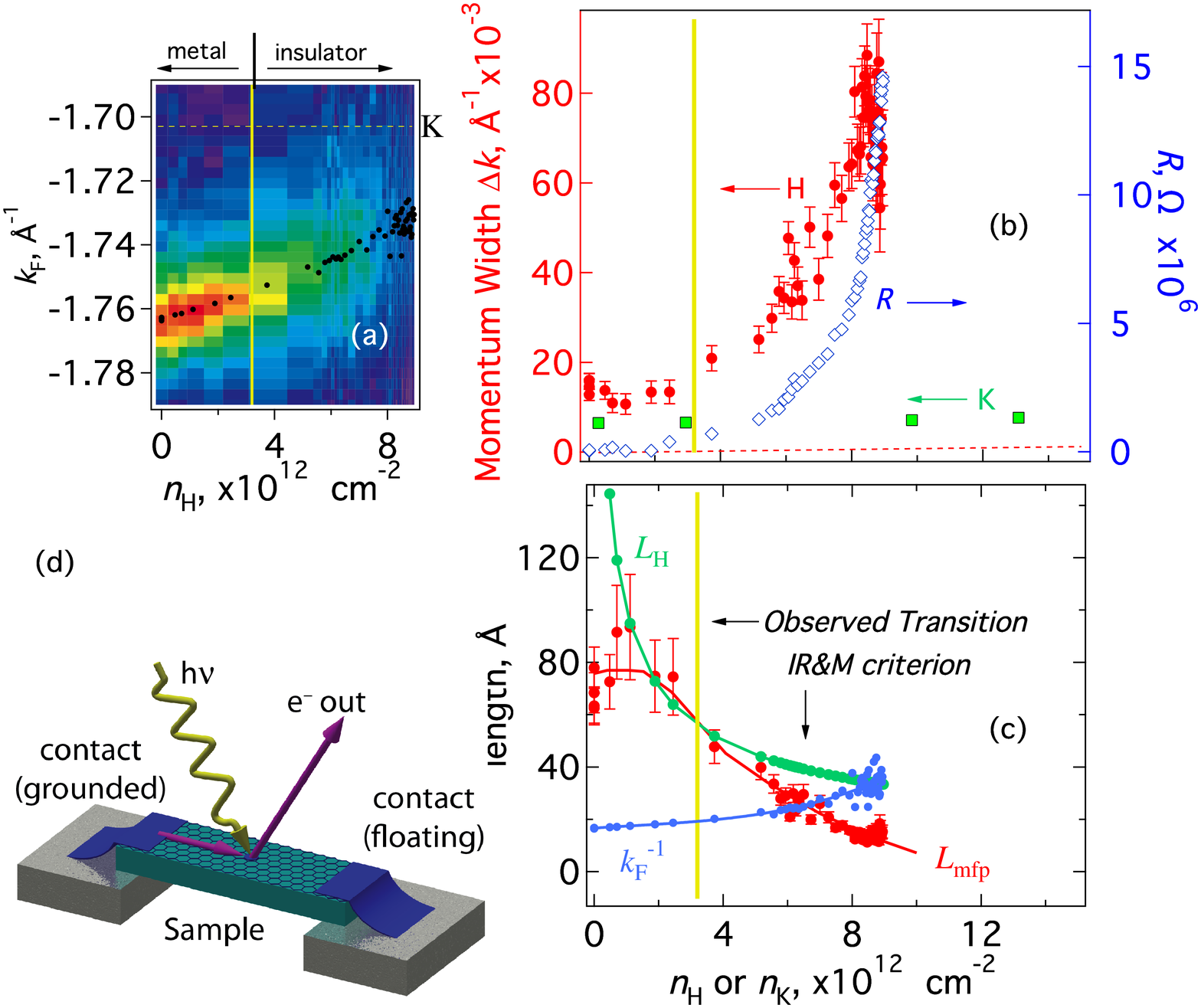}}\end{center}\caption{\figthreecaption} 
\label{fig:3}
\end{figure}
}
\def\figthreeref{Fig.\ \ref{fig:3}}

\def\figfourcaption{(a) The I-V characteristics at a single spot on the sample.  The slope of the $I$-$V$ curve defines the sample
resistance $R$. (b) The resistance plotted logarithmically against $(1/T)^{1/3}$, which should be a straight line in the VRH
model.}
\def\figfour{\begin{figure} {\includegraphics[width=6.25in]{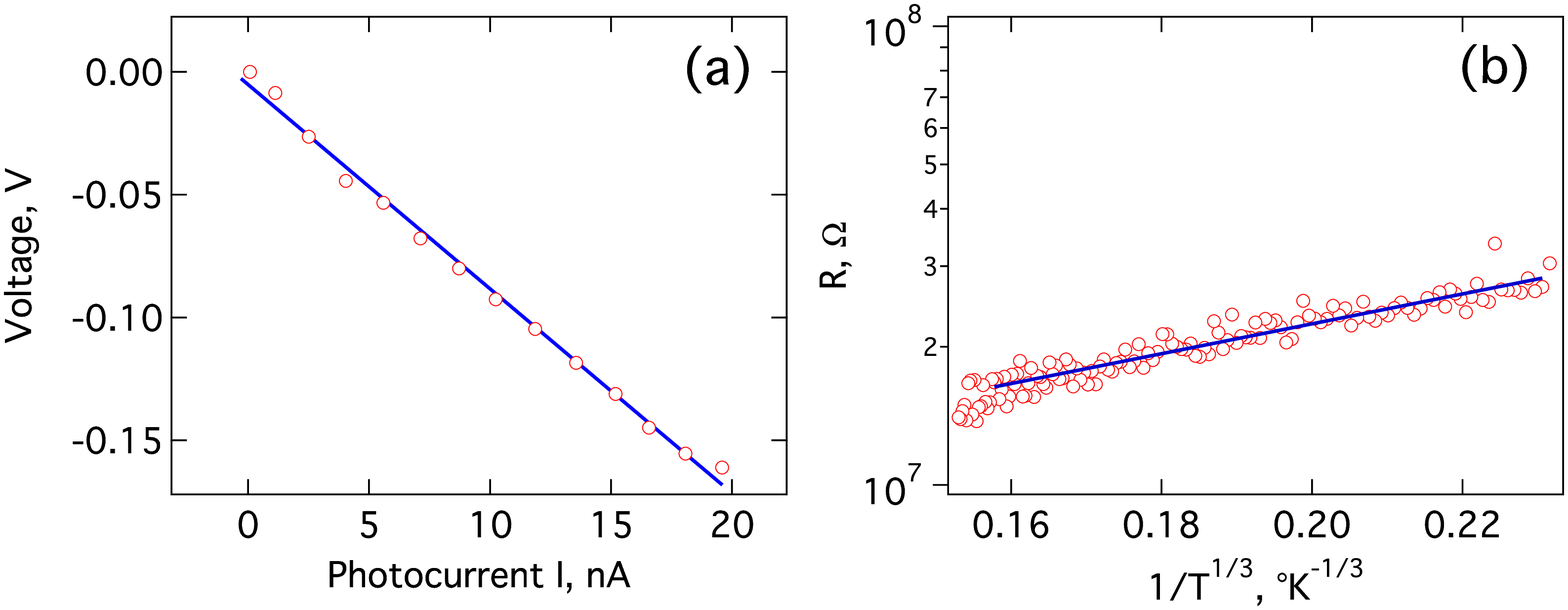}}\caption{\figfourcaption} 
\label{fig:4}
\end{figure}
}
\def\figfourref{Fig.\ \ref{fig:4}}

\begin{abstract}

Here we show, with simultaneous transport and photoemission measurements, that the graphene-terminated SiC(0001) surface undergoes a
metal-insulator transition (MIT) upon dosing with small amounts of atomic hydrogen.  We find the room temperature resistance
increases by about 4 orders of magnitude, a transition accompanied by anomalies in the momentum-resolved spectral function including
a non-Fermi Liquid behaviour and a breakdown of the quasiparticle picture.  These effects are discussed in terms of a possible
transition to a strongly (Anderson) localized ground state.
\end{abstract}

\pacs{71.30.+h,79.60.-i,73.61.Wp,72.15.Rn}
\maketitle

The interaction of atomic hydrogen with graphene is of fundamental interest.  An H atom is expected to saturate a single C \pz\
orbital, forming a local \spthree\ atomic arrangement \cite{miura2002}, which should significantly alter the local electronic structure due to the
breaking of symmetry within the 2-atom basis set of the honeycomb unit cell.  At high coverages of H
on graphene, a band-insulating behaviour has been predicted\cite{sofo2007} and possibly observed\cite{elias2008}.  
On the other hand, at low coverages H or other point defects in graphene are predicted to lead to either
magnetism\cite{duplock2004} or a localized insulating state\cite{suzuura2002,suzuura2002b,naumis2007,amini2008,robinson2008} depending on whether
the H is arranged orderly or not.  \figone

Experimentally, the low-coverage dosing of atomic H on graphite leads to either isolated H monomers or dimers,
depending on the dosing rate and temperature\cite{hornekaer2006}.  In this work, we prepare samples in the low-coverage,
low-temperature monomer regime\cite{hornekaer2006}, with areal hydrogen density \nh\ $\sim 10^{12}$ cm$^{-2}$, or around 0.1\%
coverage.  While the clean graphene samples are metallic\cite{emtsev2009} with ordinary Fermi-liquid behaviour\cite{bostwick2007},
we find, through a unique combination of transport and angle-resolved photoemission spectroscopy (ARPES), that this behaviour breaks
down at a critical coverage \nh $\sim 3\times 10^{12}$ whereupon the samples become strongly insulating, with room
temperature resistance increasing by about 4 orders of magnitude.

Pristine metallic graphene samples were prepared either using the UHV annealling method\cite{forbeaux1998,bostwick2007} or by high
pressure Ar-annealling \cite{emtsev2009}.  While the data presented are for the UHV-prepared samples, the effects observed were
similar for either method.  These samples were exposed to atomic H decomposed from H$_{2}$ gas passed through a tungsten capillary
heated to about $2000^{\circ}$C. The H pressure at the sample surface (held at $T=70^{\circ}$ K) is estimated to be $3\times 10^{-9}$ T. ARPES
measurements were conducted at the Electronic Structure Factory endstation at beamline 7 of the Advanced Light Source.

Fig.\ 1 shows the Fermi surface and bandstructure cuts of graphene as a function of H coverage.  We find that the Fermi surface area
decreases with H coverage, showing first that atomic hydrogen acts as an acceptor, and second, providing a method to determine the
coverage \nh: with the reasonable assumption of one hole per H atom, then \nh\ is given by a geometrical factor times the change in 
Fermi surface area.

Quantitative analysis of graphene's energy bands as in Ref.\ \cite{bostwick2007} shows striking trends with \nh.
Unlike ordinary chemical doping with potassium or NO$_{2}$ \cite{bostwick2007,zhou2008b}, the bands do not shift rigidly with \nh,
but instead the Fermi surface area is reduced through a curious upturn of the upper \pistar\ band above \ED, with little change to
the $\pi$ bands below \ED\ apart from a general broadening.  This upturn of the bands implies that the density of states \dw\ is
greatly reduced above \ED, see Fig.\ 2a \cite{note1}.  Such changes to the density of states  depend on the details
of the local electronic states around defects \cite{pereira2008}.

Moreover, the inverse mean free path (i.e. the momentum width \dk\ of the bands), increases far faster for the states above \ED\
than below (Fig.\ 2b).  At low \nh, the lifetime is longest near the Fermi energy \EF\ (at $\omega=0$), a sign of graphene's
ordinary Fermi liquid (FL) behaviour\cite{bostwick2007}.  But at higher \nh, the lifetime is shortest near \EF, inconsistent
with a FL.  Consequently, there is a rapid increase in the energy width $\Delta\omega$ of the quasiparticle peaks near \EF,
as shown in Fig.\ 2c.

In the quasiparticle approximation, scattering effects are sufficiently weak so as to preserve the freely propagating carriers.  But
 for sufficiently high \nh, we find $\Delta\omega/\omega>>1$ as $\omega\rightarrow 0$.  This implies a complete breakdown of the
quasiparticle picture, and evidently the states cannot be described as propagative above \ED\ for higher \nh.

As a result of the rearrangment of spectral weight, the energy spectrum at the K point, which begins as a single peak in the
pristine samples, splits with \nh\ into two peaks whose overall weight moves towards \EF\ (see Fig.\ 2d).  This latter change does
not reflect an energy gap, but merely the rearrangement of spectral weight as the \pistar\ band moves
towards the K point while simultaneously broadening.  We have previously seen quite similar changes for underannealled, but
hydrogen-free, graphene samples, and we ascribed these changes to unknown defects most likely within the bulk
graphene\cite{rotenberg2008}.  Indeed the spectra for \nh$\leq 3 \times 10^{12} cm^{-2}$ in Figs.\ 1-2 are practically the same as
in Ref.  \cite{rotenberg2008}.  This shows that the results we have found here may be generic to other kinds of point defects, not
just those induced by H atoms.

We emphasize that ARPES data are hardly sensitive to any grain boundaries or step edges, which comprise a negligible part of the
sample.  And, we find no qualitative difference for H on nominally grain-free samples \cite{emtsev2009}. Therefore,  the spectral
features arise not from any boundary edge effects but are intrinsic to the bulk graphene.  Since the same spectral features (a split
energy distribution at the K point and a reduced Fermi surface area) are seen in islanded graphene samples
\cite{rotenberg2008,zhou2008,zhou2008c}, any interpretation of such spectral features in terms of island geometry or edge states
would be debatable.

Fundamentally, the asymmetry with respect to \ED\  arises from the position of a dispersionless hydrogen acceptor level,
which is not centered at \ED\ as for carbon vacancies (in a treatment without second neighbor hopping\cite{pereira2008}) but instead 
$\sim 200$ meV above \ED\cite{duplock2004,robinson2008,lovvik2008}.  This state, although weak, is observed by comparing ARPES for the clean and disordered
sample obtained far from the $\pi$ bands (Fig.\ 2e); it imposes another particle-hole symmetry breaking factor discussed later. 

Now we discuss the dosing dependence of the states near \EF\ in more detail.  Fig.\ 3a shows the momentum distribution of the states at
\EF\ \vs\ \nh.  While at all coverages, the peak of the momentum distribution ($\sim$\kF) disperses towards the K point, at a
critical coverage, the state becomes abruptly much broader, reflected in a plot of the momentum width at \EF\ in Fig.\ 3b.  Such an
abrupt change in width does not occur for potassium atoms (data also shown in Fig.\ 3b), or for binding energies below \ED\ (not
shown) and reflects an abrupt transition of the spectral function.

\figtwo
\figthree

Furthermore, we observe a dramatic increase of sample resistance at this coverage, also shown in Fig.\ 3b, measured as follows: in
the photoemission experiment, a compensating current ($I\sim 10$ nA) travels from one side of the sample to the photoemission site
(see Fig.\ 3d).  A resistance $R\gtrsim$ 0.2 M$\Omega$\ along this path induces a measurable voltage shift $V=IR$ of the spectrum.   Furthermore, by measuring the differential
resistance at two spots, we could not only prove that the resistance arose within the graphene (and not, e.g. at the sample clip),
but also accomplish the equivalent of a 4-point resistance measurement free of any contacts.

By varying the incident photon flux, we could measure a linear $I$-$V$ relationship, demonstrating Ohmic resistance, see Fig.  4a.
More importantly, the temperature-dependent resistance $R(T)$ (Fig.  4b) displays insulating behaviour (i.e. $dR/dT<0$), thus ruling
out that the sample is a ``bad metal.''  Since the pristine samples are metallic ($dR/dT<0$)\cite{emtsev2009}, we must conclude that
the breakdown of the Fermi liquid and other anomalous spectral features occur at the same time as a metal-insulator transition
(MIT).

What is the fundamental cause of the MIT? The reconstruction of the energy spectrum by H atoms, although dramatic, preserves the
Fermi surface, and therefore formation of an ordinary band insulator such as graphane\cite{sofo2007} is not supported by the data.
We can similarly rule out other gap-inducing ground states such as density waves or Mott insulating behaviour.

A similar defect density causes an MIT to an Anderson localization (AL) state in carbon nanotubes\cite{gomez2005,flores2008}.  So it is
tempting to ask whether we have induced an AL insulating state in graphene.  Such a transition was predicted for graphene, depending
on the symmetry class, density and range of the defects\cite{suzuura2002,suzuura2002b,aleiner2006,robinson2008} and sample geometry\cite{lherbier2008}.
While long-range impurities preserve the local lattice symmetry and suppress backscattering, leading to
anti-localization\cite{wu2007}, short-range impurities (those that, like H, break the symmetry of the graphene unit cell) lead to
intervalley scattering, a significant backscattering, and, as a result, weak\cite{morozov2006} or strong localization (i.e. AL).

Tight-binding calculations for graphene with on-site disorder show that a significant mobility gap ($\sim400$ meV) in the extended
states can be developed for a modest density of short-range scatterers ($\sim 0.1$\%), comparable to our dosing level
\cite{naumis2007,amini2008}.  Such a mobility gap, if centered on the H acceptor level we observed, would reach \EF\ and induce AL there.   
Renormalization group calculations for H on graphene also support the appearance of zero conductivity for similar coverages to what 
we observe \cite{robinson2008}.

The temperature-dependent resistance $R(T)$ is consistent with AL since log $R$ is linear with $(1/T)^{1/3}$ (Fig.\ 4b), following
Mott's well-known relationship for a variable-range-hopping (VRH) model in 2D\cite{mott1979}.  VRH is
necessary as evidence for AL, but it cannot be regarded as sufficient proof in this case: First, the accessible temperature
range did not give enough change in resistance for a conclusive analysis of the exponent; second, and more fundamentally, VRH
transport does not occur only for AL, but also in any system with localized states such as disordered insulators or granular media.

However, the \emph{simultaneous} measurement of ARPES with transport as a function of defect density not only rules out these
other mechanisms, but the observed breakdown of the quasiparticle picture provides positive evidence for the AL transition in our
samples.  Since localized states are not quasiparticles, we expect a fundamental change in the spectral function upon
localization\cite{mattis1973}, as we have observed.  Whereas ordinarily the momentum width at \EF\ reflects the mean free path of the quasiparticles,
we propose that upon localization it should instead reflect the localization length of the electrons.

The linewidth analysis of the ARPES data (Fig.\ 3c) supports this picture.  Here we plot the mean free path \lmfp\ (i.e. the inverse
momentum width at \EF) \vs\ defect density \nh, and compare it to a hydrogen length scale given by \lh$=1/\sqrt{n_{\mathrm{H}}}$.
Above the critical coverage, \lmfp\ is limited by the distance between hydrogen atoms, as expected in the above picture.  The MIT
occurs at a similar, but lower defect density than the simple prediction by Ioffe, Regel, and Mott(IR\&M), which predicts the MIT when
\lmfp$\sim 1/$\kf \cite{ioffe1960,mott1979}, or, in our case, when \nh $\sim 6\times 10^{12}$ cm$^{-2}$.

The presence of a dispersionless H state\cite{duplock2004,lovvik2008,robinson2008} between \ED\ and \EF\ is of fundamental importance  since it
provides a microscopic model for how AL can take place.  It explains the preferential broadening of the \pistar\ band due to an
energy-dependent scattering cross-section \cite{basko2008,robinson2008}, and it suggests that the energy distribution of localized states is not
centered on \ED, as expected from the diverging de Broglie wavelength (and predicted within particle-hole-symmetric
models\cite{naumis2007,amini2008}) but is instead biased to the \pistar\ states near \EF\cite{robinson2008}.

Some subtle changes to the spectral functions in Fig.\ 1 can now be explained within this model.  The kinks and associated linewidth
variations in the bands due to electron-phonon interaction, studied in detail for metallic graphene \cite{bostwick2007,
mcchesney2008} are clearly preserved in the metallic state, but become washed out in the insulating state, another clear breakdown of the
quasiparticle picture.  In contrast, defects normally show up as an additional, uncorrelated scattering term in the electronic
self-energy and do not much affect the electron-phonon kinks\cite{hengsberger1999b}.  Intriguingly, the kink disappears not through some
overall reduction in its strength, but rather by an apparent reduction in its energy scale, i.e. the band kinks at an energy
progressively closer to \EF. This may reflect the motion of an underlying mobility edge in the extended states towards \EF.

\figfour

A recent ARPES study of graphene claimed to induce an MIT by lowering \EF\ into a pseudogap near \ED\cite{zhou2008b}.  In a recent
transport study\cite{elias2008}, a much higher H dose than we used ($\sim 50\% $ coverage) was claimed to create a disordered
insulator state, characterized by clustering of H atoms correlated to ripples.  Both of these systems were described as ordinary
insulators created by chemical means, although with ambiguity about the actual resistance in the first study \cite{zhou2008b}, and
about the bandstructure, density and distribution of defects in the second \cite{elias2008}.  In contrast, our finding of an
insulating transport coexisting with an intact Fermi surface is inconsistent with such a band insulator in our case and provides
tantalizing evidence in support of the AL ground state.

This conclusion if true suggests promising directions for future research.  Access to the momentum-dependent spectral function can open a new
experimental window onto the role and interplay of interactions with the Anderson localization transition.  While we have studied
randomly placed defects, one can envision a patterned deposition of H atoms, or even the atomically-controlled placement of defects
using scanning probe manipulation, so that many important aspects of localization theory could be studied systematically.  As an
easily prepared material, accessible not only to transport and gating but also a variety of surface probes, our results suggest
that graphene is a new paradigm for resolving open questions about conductance in disordered materials.

\acknowledgments This work was supported by the Director, Office of Science, Office of Basic Energy Sciences, of the U.S, Department
of Energy.


\begin{thebibliography}{32}
\bibitem[{not()}]{note1}
\bibinfo{note}{The two dimensional density of states is derived from the
  experimental band dispersion $k(\omega)$ through
  $\mathcal{D}(\omega)=(2/\pi)k(\omega)dk(\omega)/d\omega$}.
\end{thebibliography}

\end{document}